\documentclass[aip,graphicx, reprint]{revtex4-1}

\draft 

\usepackage{lineno}
\usepackage{xcolor}
\usepackage{graphicx}

\graphicspath{{../figures/}}

\draft 

\begin{document}

{\color{blue}
{\scriptsize\textit{ The following article has been accepted by Journal of Applied Physics. After it is published, it will be found at \href{https://aip.scitation.org/journal/jap}{Link}.}} }

\title{High conductivity Polarization-induced 2D hole gases in Undoped GaN/AlN Heterojunctions enabled by Impurity Blocking Layers} 

\author{Reet Chaudhuri}
\email[]{rtc77@cornell.edu}
\affiliation{School of Electrical and Computer Engineering, Cornell University}

\author{Zhen Chen}
\affiliation{Department of Applied and Engineering Physics, Cornell University}

\author{David Muller}
\affiliation{Department of Applied and Engineering Physics, Cornell University}
\affiliation{Kavli Institute of Nanoscience, Cornell University}

\author{Huili Grace Xing}
\affiliation{School of Electrical and Computer Engineering, Cornell University}
\affiliation{Department of Material Science and Engineering, Cornell University}
\affiliation{Kavli Institute of Nanoscience, Cornell University}

\author{Debdeep Jena}
\affiliation{School of Electrical and Computer Engineering, Cornell University}
\affiliation{Department of Material Science and Engineering, Cornell University}
\affiliation{Kavli Institute of Nanoscience, Cornell University}

\date{\today}


\begin{abstract}
High-conductivity undoped GaN/AlN 2D hole gases (2DHGs), the p-type dual of the AlGaN/GaN 2D electron gases (2DEGs), have offered valuable insights into hole transport in GaN and enabled the first GaN GHz RF p-channel FETs. They are an important step towards high-speed and high-power complementary electronics with wide-bandgap semiconductors. These technologically and scientifically relevant 2D hole gases are perceived to be not as robust as the 2DEGs because structurally similar heterostructures exhibit wide variations of the hole density over $\Delta p_s > 7 \times 10^{13}$ cm$^{-2}$, and low mobilities. In this work, we uncover that the variations are tied to undesired dopant impurities such as Silicon and Oxygen floating up from the nucleation interface. By introducing impurity blocking layers (IBLs) in the AlN buffer layer, we eliminate the variability in 2D hole gas densities and transport properties, resulting in a much tighter-control over the 2DHG density variations to $\Delta p_s \leq 1 \times 10^{13}$ cm$^{-2}$ across growths, and a 3$\times$ boost in the Hall mobilities.  These changes result in a 2-3$\times$ increase in hole conductivity when compared to GaN/AlN structures without IBLs.
\end{abstract}


\maketitle 


\section{Introduction}
The primary p-type acceptor dopant for gallium nitride (GaN) since 1990s has been Magnesium (Mg) with a thermal activation energy of $\sim170$ meV \cite{Nakamura2015, Kozodoy2000a}. The low mobile hole activation of $<$4\% at room temperatures combined with acoustic phonon scattering limited hole mobilities\cite{Ponce2019} of $<50$ cm$^2$/Vs in GaN at room temperature result in $\sim 10 - 100\times$ higher resistivity in p-GaN layers compared to n-GaN layers of comparable thickness. Polarization-induced doping provides an alternate route to generate holes in III-nitride heterostructures by taking advantage of the spontaneous and piezoelectric polarization discontinuities. The recent observation \cite{Chaudhuri2019} of a high-density 2D hole gas (2DHG) in undoped GaN/AlN structures experimentally confirmed that this technique\cite{Ambacher1999} generates holes at a negative polarization-difference interface without the need of acceptor doping.
 
In these structures, when strained metal-polar GaN is grown on AlN, a \textit{negative} fixed sheet charge of $\sigma_{\pi} \sim5\times 10^{13}$ cm$^{-2}$ is induced at the GaN/AlN interface \cite{Ambacher1999}. If the GaN layer is thicker than a critical value of $\sim3$ nm, this fixed sheet charge generates an enormous electric field that field-ionizes electrons out of the valence band states to localized surface states, leaving behind mobile \textit{holes}. These holes are confined in 2D in a triangular quantum well at the heterojunction formed by the valence band (VB) offset between GaN and AlN. The formation of such polarization-induced 2D hole gases in principle should not require acceptor dopants in a carefully designed and grown GaN/AlN heterostructure. 

But in the past, 2DHGs in undoped GaN/AlN structures have not been as robust as their n-type analog, the undoped AlGaN/GaN 2D electron gases (2DEGs). In structurally similar GaN/AlN samples, mobile 2DHGs were either absent altogether, or if present, had widely varying sheet densities and mobilities.  Mg doping in the top GaN layer \cite{Li2013} was found to stabilize the 2DHG and demonstrate p-channel FETs using such 2DHGs. Only recently were these variations solved to demonstrate the high-conductivity 2DHGs in undoped GaN/AlN structures\cite{Chaudhuri2019}. In this report we identify the reason for the earlier variations, and demonstrate a method to obtain these repeatable high-conductivity 2DHGs with a tighter control over densities and transport properties. 
 
Using GaN/AlN 2DHGs as the channel, scaled GaN p-channel heterostructure field effect transistors (p-HFETs) with record high on-currents exceeding $400$ mA/mm were reported recently\cite{nomotoIEDM2020} which, for the first time, broke the GHz speed barrier with cut-off frequencies in the 20 GHz regime. With key p-channel FET device parameters making a climb towards that of GaN n-channel HEMTs, this result represents a significant step towards enabling high-voltage RF wide-bandgap complementary device platforms \cite{Bader2020ProspectsDevices}. The high polarization-induced 2DHG densities are crucial for achieving low contact and access resistances and the resulting high on-currents in these p-HFETs. Variation in 2DHG density directly translates to variations in the transistor threshold voltage and drive current that are undesirable. Repeatable growth of these 2DHGs with tight control on the density and high conductivity is therefore critical.  

In this work we find that silicon (Si) and oxygen (O) impurities originate from the starting substrate and float up and incorporate in the GaN layer at the GaN/AlN heterojunction. Because both are donors in GaN, they partially compensate the 2DHG and act as scatterers of mobile holes, resulting in inconsistent densities, lower mobilities and higher sheet resistances.  We present a new approach of impurity blocking layers (IBLs) buried in the underlying AlN. The IBLs effectively prevent the impurities from reaching the GaN/AlN interface where the holes are located, significantly enhancing the repeatability and control over the densities, mobilities, and conductivity of the undoped 2DHGs.

\begin{table*}[ht!]
\caption{\label{tab1}Summary of the samples under study and their Hall-effect measurement results. Samples A, B and C were conductive at 300 K and 77 K with a positive Hall coefficient and increased mobility at low temperatures. The control sample with only AlN buffer layer was highly resistive. Therefore the measured conduction in samples A, B and C are expected to arise from the polarization-induced 2DHG at the GaN/AlN interface.}
\begin{tabular}{ | c | c | c | c | c | c | c | }
\hline
 ID & Structure &
  Technique Used in AlN Buffer Layer &
  \begin{tabular}[c]{@{}c@{}}{}300K $p_s$ \\ (10$^{13}$ cm$^{-2}$)\end{tabular} &
  \begin{tabular}[c]{@{}c@{}}{}300K R$_{\textrm{sheet}}$ \\ $\Omega$/sq\end{tabular} &
  \begin{tabular}[c]{@{}c@{}}{}$\mu_\textrm{{300K}}$ \\ ( cm$^2$/Vs)\end{tabular} &
  \begin{tabular}[c]{@{}c@{}}{}$\mu_\textrm{{77K}}$ \\ ( cm$^2$/Vs)\end{tabular} \\ 
  \hline
  \hline
Sample A & 15 nm GaN/420 nm AlN & N-rich nucleation layer (NL)    &  5.48  & 22.4$\times10^{3}$ & 5.11 & 19.1\\ 
Sample B & 15 nm GaN/430 nm AlN & Impurity Blocking Layers (IBLs) & 4.86  & 8.67$\times10^{3}$ & 14.8 & 54.2\\ 
Sample C & 15 nm GaN/440 nm AlN & N-rich NL + IBLs                & 5.24 & 11.78$\times10^{3}$ & 10.1 & 48\\ \hline
IBLs only & 440 nm AlN & N-rich NL + IBLs                & - & $>2\times10^{6}$ & - & - \\ \hline
\end{tabular}
\end{table*}
 
 \section{Experiment}
 
The heterostructures studied here were grown by plasma-assisted molecular beam epitaxy (MBE).  The active nitrogen (N) species is supplied from a RF plasma source. Gallium (Ga) and aluminum (Al) metal fluxes were supplied from effusion cells. The starting substrate surface contains chemical impurities such as Si, O, H, C from exposure to atmosphere. For layer-by-layer epitaxial growth of AlN or GaN, films are grown under metal-rich condition\cite{Koblmueller2003, Heying2000ControlEpitaxy}, i.e. the incident metal to nitrogen flux ratio is $>1$. When AlN is nucleated on a substrate in Al-rich conditions, the Si impurities present on the surface are thermodynamically inhibited from incorporating into the crystal. The relative thermodynamic formation energies of the resultant species during growth \cite{Hoke2014} dictates that the preference of incorporation of competing cation species into the crystal is Al $>$ Si $>$ Ga $>$ In. Thus, the Si atoms must either desorb, or float up on the Al metal adlayer while the AlN layer is grown. Typical MBE growth temperatures of 800$^{\circ}$C-1000$^{\circ}$C necessary for smooth epitaxy is insufficient to desorb Si from the AlN surface, implying the Si must float on the Al adlayer. 

When a GaN layer is grown on top of the AlN layer, it now becomes thermodynamically favorable for the Si which floated up to incorporate\cite{Hoke2005, Poblenz2005}. For GaN/AlN heterostructures where the mobile holes are expected at the interface, this Si incorporation poses a significant problem.  In GaN RF HEMTs grown on SiC substrates, the initial AlN nucleation layers are typically grown in the nitrogen-rich (N-rich) condition (Al:N flux ratio $<1$) to suppress Si out-diffusion from a SiC nucleation interface. This forces the Si to incorporate into the N-rich AlN nucleation layer (NL)\cite{Hoke2005, Poblenz2005}. However N-rich MBE growth condition also leads to incomplete coverage and structural defect densities in the film \cite{Heying2000ControlEpitaxy, Koblmueller2003} which are undesirable for 2DHG transport. 

High conductivity 2DHGs with tight density control should then be possible by introducing impurity blocking layers (IBLs). The thermodynamic preference of incorporation\cite{Iliopoulos2011GrowthEpitaxy, Bottcher2012} dictates that though Si cannot incorporate in metal-rich AlN, it readily incorporates in AlGaN layers by substituting the Ga sites. For a constant dopant flux the Si incorporation in AlGaN shows almost no change with Al\% in the crystal till 100\% Al content (ie. AlN), is reached\cite{Lee2020ImprovingLasers}, where the Si incorporation abruptly drops to zero. Thus ultrathin high Al-composition AlGaN layers inserted in the AlN should effectively suppress the Si up-diffusion from the substrate. Furthermore, high-composition $>90\%$ and thin $<1-2$ nm AlGaN layers will allow the layers to remain fully-strained to the AlN substrate without introducing additional structural defects. The thin layers also prevent parallel conduction channels in the form of 2DEGs due to polarization difference at the AlGaN/AlN interfaces.  The small band offsets make the bound states in the quantum well shallow. These thin, high-composition $>90\%$ AlGaN layers IBLs are investigated in this study in comparison to the N-rich nucleation layers mentioned above.  The effect of such IBLs on the chemical, structural and transport properties of the UID-GaN/AlN 2DHG heterostructures is next described. 

\textbf{Table \ref{tab1}} shows a series of samples that were grown using a Veeco Gen10 MBE system under nominally similar conditions. 8 mm $\times$ 8 mm diced wafers of metal-polar MOCVD-grown 1 $\mu$m AlN on sapphire templates were used as starting substrates. The substrates were ultrasonicated in acetone, methanol and iso-propanol for 15 minutes each and mounted on lapped 3-inch Si wafers using molten indium. They were then out-gassed overnight at 200$^{\circ}$C for 7 hours, at 500$^{\circ}$C for 2 hours in vacuum before introducing into the MBE growth chamber. Magnesium and silicon dopant cells were kept cold ensuring all films were undoped, or unintentionally-doped (UID). The RF plasma power was maintained at 400 W corresponding to a N-limited growth rate of $0.52 \ \mu$m/hour. All samples included a $\sim420-440$ nm AlN buffer layer grown at a substrate temperature T$_{\text{sub}}$ = 780$^{\circ}$C, followed by a 15 nm layer of undoped GaN grown at T$_{\text{sub}}$ = 750$^{\circ}$C. The three samples differed structurally with respect to the type of the AlN buffer layer. 

These differences are qualitatively highlighted below, and described in \textbf{Table \ref{tab1}}: 

\begin{itemize}
    \item In sample A, the first $\sim$30 nm of AlN nucleation layers (NL) was grown under N-rich conditions with Al:N flux ratio $\sim$0.9. The rest of the buffer layer was grown under metal-rich condition with Al:N flux ratio $\sim$1.33, for a total thickness of $\sim$420 nm. No IBLs were introduced in the buffer layer in this sample.

    \item In sample B, the AlN was grown under Al-rich conditions throughout. Thin IBLs with high-composition Al were incorporated periodically every $\sim$55 nm in the AlN buffer layer. A total $\sim$430 nm thick AlN buffer layer was grown with 8 IBLs.

    \item In sample C, the AlN buffer layer with $\sim$30 nm of N-rich AlN nucleation layer followed by Al-rich AlN with 5 IBLs separated by $\sim$55 nm was grown, for a total thickness of $\sim$440 nm. The layer structure is shown in \textbf{figure \ref{f1}(a)}.
    
    \item A control sample with only AlN with IBLs and no GaN to investigate the background resistivity.
\end{itemize}

After the $\sim420-440$ nm AlN buffer layer growth, the substrate was cooled down to T$_{\text{sub}}$ = 750$^{\circ}$C for the growth of the UID GaN layer. The 2DHG is expected to form at this GaN/AlN interface. The thickness of the UID GaN layers was $\sim$15 nm for all three samples, grown under the same nominal metal-rich conditions with Ga BEP flux of $\sim1\times 10^{-6}$ Torr. Detailed MBE growth recipes are provided in the supplementary section.

 \section{Results and Discussions}

\begin{figure}[ht!]
\begin{center}
\includegraphics[keepaspectratio=true, width=0.475\textwidth]{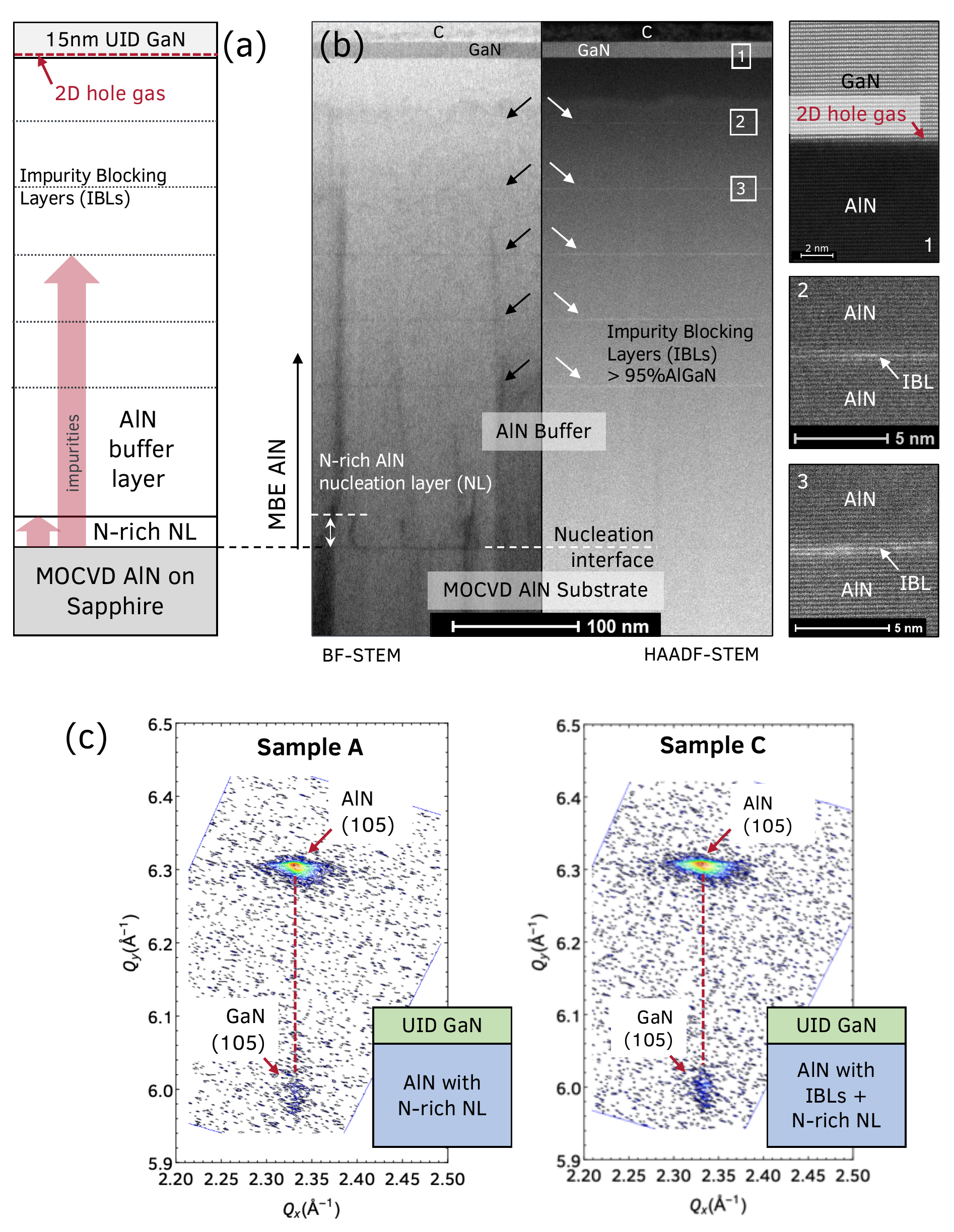}
\end{center}
\caption{\label{f1} (a) Heterostructure details of sample C, with an AlN nucleation layer (NL) grown under N-rich conditions and impurity blocking layers (IBLs) marked in the buffer layer. (b) Scanning transmission electron microscope (STEM) images of the cross-section of a GaN/AlN heterostructure of sample C. The bright-field (BF) image shows the presence of the N-rich NL. The IBLs are visible in both BF and high angle annular dark field (HAADF) image. (c) X-ray diffraction Reciprocal space maps (RSM) around the AlN (105) peak confirms that the GaN layers in sample A and C are fully strained to the AlN buffer layer.}
\end{figure}

\textbf{Figure \ref{f1}(b)} shows STEM scans of the cross-section of sample C, taken using a bright-field (BF) detector and a High-Angle Annular Dark Field (HAADF) detector. Threading dislocations propagating through the substrate to the surface are visible, a signature of heteroepitaxial AlN growth. The N-rich nucleation layer is visible in the BF scan, and shows contrast due to the slightly higher dislocation density in the layer. Most of these dislocations, originating from the N-rich nucleation interface, do not propagate beyond the N-rich layer. The scans from the HAADF detector, which detect the atomic number $Z$ contrast clearly show the periodic high-composition AlGaN IBL layers located $\sim55$ nm apart. The zoomed-in scan of the IBLs shows the Ga incorporation in the AlN over 2-3 monolayers (MLs). The GaN /AlN interface where the 2D hole gas is expected is also shown, which is smooth and abrupt as desired.  

The thin GaN layers are expected to be compressively strained to the AlN buffer layer due to the 2.4\% in-plane lattice mismatch. X-Ray diffraction reciprocal space maps (RSMs) shown in \textbf{figure \ref{f1}(c)} confirm this. The relative position of the AlN and GaN (105) peaks show that the GaN layers are pseudomorphically strained to AlN buffer layer in both sample A and C. Only the AlN and GaN diffraction peaks are visible with no discernible peak from the AlGaN IBL layers due to their composition and thinness. The RSMs confirm there was no structural effect of the IBLs on the GaN layer, and that the samples are similar structurally and as targeted for generating a 2DHG. Omega rocking curve scans taken around the GaN (002) peak (not shown here) confirm that all three samples have similar screw-type threading dislocation densities of $\sim10^8$ cm$^{-2}$.

Results of room temperature Hall-effect measurements, performed at magnetic field of 0.32 T using soldered corner Indium contacts to the 2DHG, are summarized in \textbf{Table \ref{tab1}}. The three samples with the undoped GaN epitaxial layer showed a repeatable positive Hall coefficient confirming the presence of mobile holes. The control sample with IBLs only was found to be highly resistive, confirming that the IBLs in the buffer layer do not form parallel conduction channels. Since no p-type acceptor doping is present in the sample, the conduction in samples A, B and C are entirely due to the 2DHG at the GaN/AlN interface. The fact that it indeed is a 2DHG is confirmed by the fact that the density does not freeze out at 77 K, and the mobility increases.  Before discussing the hole transport properties further, we report the chemical effect of the IBLs.

\begin{figure*}[htb!]
\begin{center}
\includegraphics[keepaspectratio=true, width=0.9\textwidth]{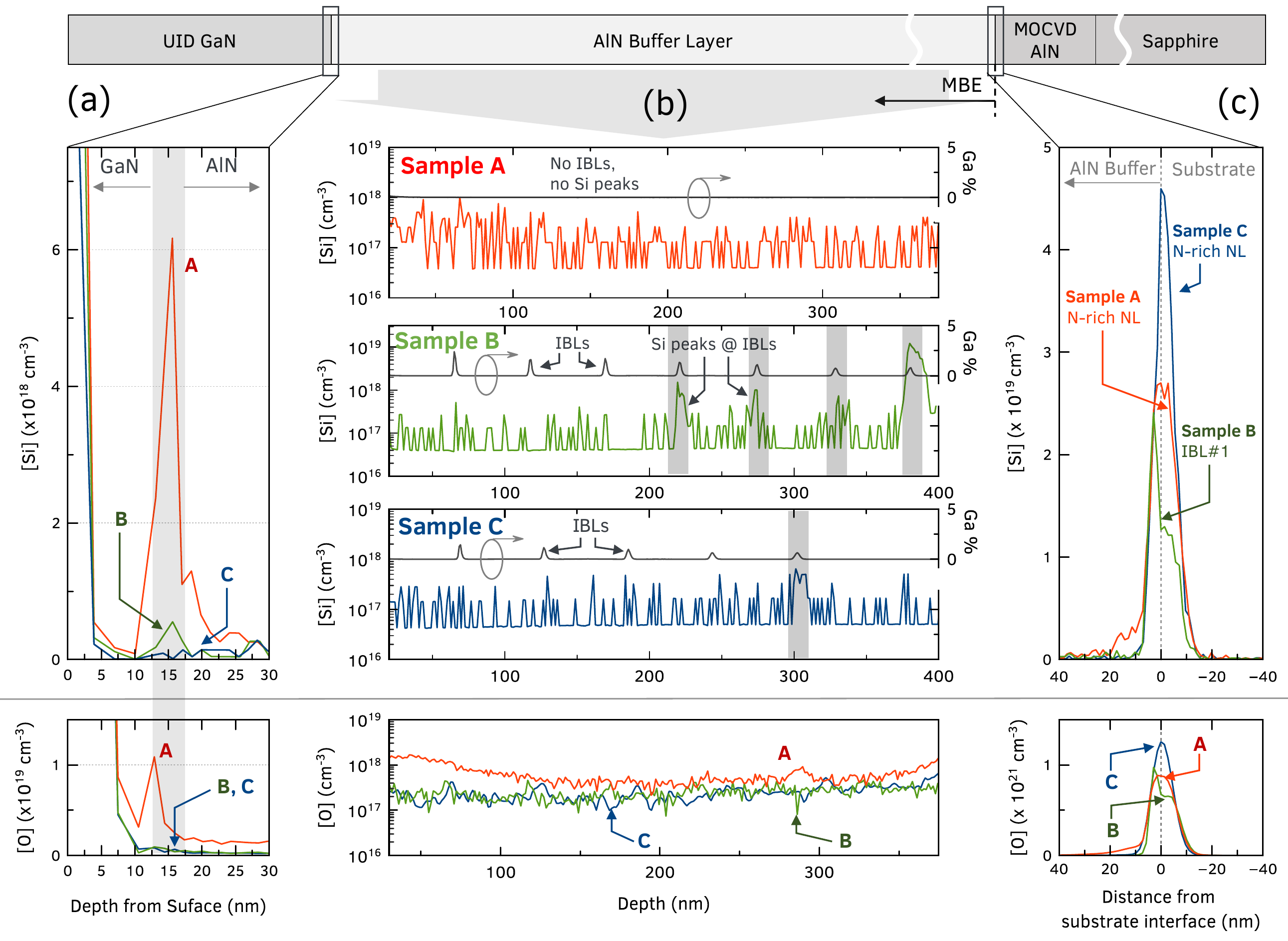}
\end{center}
\caption{\label{f2} Secondary ion mass spectrometery (SIMS) profiles showing the impurity levels and Ga atomic fractions : (a) at the GaN-AlN interface where the 2DHG is expected, (b) in the AlN buffer layer, and (c) at the AlN-substrate nucleation interface, for the three samples A, B and C. The positions of the IBLs are marked by the peaks in the Ga atomic compositions in (b). The Si peaks corresponding to the IBLs in sample B and C are also indicated. The results show that the high-concentration AlGaN IBLs are effective in blocking the impurities from floating up and incorporating in the GaN layer.}
\end{figure*}

Secondary ion mass spectrometry (SIMS) analysis was performed on the three samples to investigate the impurity profiles of the films and the effect of the different buffer layers and IBL designs. The measured concentration profile for Si and O impurities are shown in \textbf{figure \ref{f2}}, along with the Ga  atomic fractions. The SIMS detection limits are $1-2 \times 10^{17}$ atoms/cm$^3$ for Si and O. The three regions of interest from the right side are:  (c) the AlN/substrate nucleation interface, (b) the AlN buffer layer with IBLs, and (a) the GaN-AlN interface where the 2DHG is expected.  They are plotted separately for ease of comparison.

In sample A, concentration spikes of Si and O are observed at the AlN/substrate nucleation interface. The peak concentrations are $\sim3\times 10^{19}$ cm$^{-3}$ and $\sim1\times 10^{21}$ cm$^{-3}$ respectively. The Si profile tails into the AlN buffer layer for a distance of $\sim$20 nm. No Ga is detected in the AlN buffer layer. This is expected as no IBLs were grown in Sample A. No Si, O peaks are observed in the buffer layer, with the Si and O impurity levels remaining at the detection limit of around $\sim1 \times 10^{17}$ cm$^{-3}$ and $\sim5 \times 10^{17}$ cm$^{-3}$. It is noted that the O level starts rising slowly in the top 100 nm of the AlN buffer layer. However, at the GaN/AlN interface, a clear concentration spike is observed for both Si and O profiles. The peak concentrations are $\sim6\times 10^{18}$ cm$^{-3}$ and $\sim 1\times 10^{19}$ cm$^{-3}$ respectively. This points to the fact that despite using an N-rich NL layer in the AlN, some of the Si and O floated up during the growth of the metal-rich AlN buffer layer and incorporated in the initial layers of the epitaxial GaN layer. The Si spike extends over $\sim3$ nm into the GaN layer, where the 2DHG is expected to reside. Clearly, the $\sim 30$ nm N-rich NL in this sample is not effective enough to prevent the impurities from floating up during the growth of the AlN buffer layer.

The panel labeled Sample B shows the impurity profile when IBLs are introduced. The SIMS profile shows the incorporation of Ga, each peak corresponding to an IBL. The resulting AlGaN layers have $>95\%$ Al composition, in agreement with the STEM scans in \textbf{figure \ref{f1}}. The Si concentration exhibits peaks at the nucleation interface, and subsequent peaks coincide with the position of the IBLs. The Si peak density is $\sim2\times 10^{19}$ cm$^{-3}$ for the first IBL at the nucleation interface, $\sim 1\times 10^{19}$ cm$^{-3}$ for the next and then drops to $\sim1\times 10^{18}$ cm$^{-3}$ for the next 3. No Si peak is observed after 5 IBLs, beyond which the Si concentration remains at the detection limit of $\sim1\times 10^{17}$ cm$^{-3}$ even at the IBL positions. This confirms that the Si observed at the GaN/AlN interface is not from the growth chamber ambient but is floating up from the substrate surface, as otherwise Si peaks would have occurred at every IBL position and not just the first 5. A small apparent Si spike of $\sim5\times 10^{17}$ cm$^{-3}$ in \textbf{figure \ref{f2}(a)} visible at the GaN/AlN interface is around the same magnitude of the Si concentration in the adjacent AlN buffer layer in \textbf{figure \ref{f2}(b)}. However this concentration is an order of magnitude lower than the concentration in sample A. The O concentration profile in the AlN buffer layer does not show any spikes and is at the detection limit of $\sim2\times 10^{17}$ cm$^{-3}$. There is no O peak observed at the GaN/AlN 2DHG interface. This evidence suggests that the IBLs in sample B are far more effective in blocking Si and O compared to the N-rich NL in sample A. Further investigations are underway to determine the difference in the chemical behavior of the blocking of oxygen and silicon impurities. 

The SIMS profile of sample C shows the impurity concentrations when both the the N-rich NL and IBLs are used in the AlN buffer layer. Si and O concentration spikes are observed at the AlN nucleation interface, with peak concentrations of $\sim5\times 10^{19}$ cm$^{-3}$ and $\sim 1\times 10^{21}$ cm$^{-3}$  respectively. The Ga atomic fraction profile in the AlN buffer layer that the IBLs are spaced $\sim$55 nm apart with a concentration of $>95\%$ AlGaN. A small Si peak of $\sim5\times 10^{17}$ cm$^{-3}$ is observed only at the first IBL, beyond which the Si profile remains at the detection limit of $\sim1\times10^{17}$ cm$^{-3}$. The O profile does not show any peaks in the buffer layer and remains at $\sim2\times 10^{17}$ cm$^{-3}$. Importantly no Si and O peaks are observed at the GaN-AlN interface. The impurity levels at the interface are an order lower than that seen in sample A at $\sim1\times 10^{17}$ cm$^{-3}$ and $\sim1\times 10^{18}$ cm$^{-3}$ for Si and O respectively. These results suggest that the combination of N-rich NL and IBLs used in sample C is most efficient in blocking the Si, O impurities floating up from the substrate from reaching the GaN layer where the 2DHG is located. Indeed this is borne out in the transport properties of the 2DHGs.

\begin{figure}[htb!]
\begin{center}
\includegraphics[keepaspectratio=true, width=3.4in]{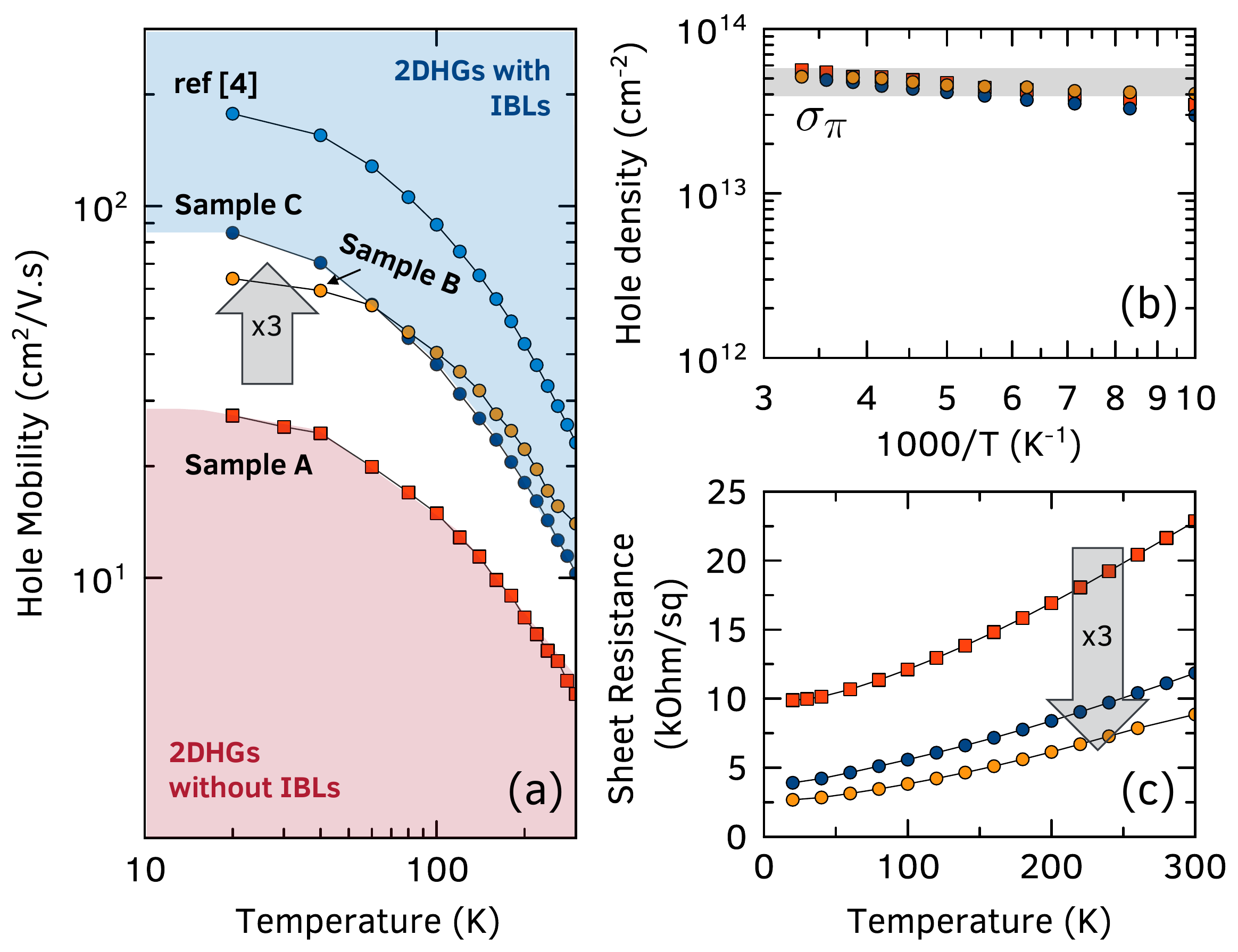}
\end{center}
\caption{\label{f3} Temperature dependent Hall-effect measurement results for sample A, B and C. (a) Comparing the holes mobilites, $3\times$ boost in mobilities is observed in sample  B and C compared to sample A throughout the temperature range due to the suppression of impurities at GaN/AlN interface. Data from other samples grown under similar conditions \cite{Chaudhuri2019} is also included, and shows high hole mobilities upto $\sim$190 cm$^2/$Vs at 20 K. (b) The 2DHG sheet concentration is approximately the same in all the samples. (c) This results in 3$\times$ higher conductivity in sample B and C compared to sample A.}
\end{figure}

The measured 300 K and 77 K Hall-effect data for the three samples A, B and C is presented in \textbf{Table \ref{tab1}}. Temperature-dependent Hall-effect measurements were performed compare the 2DHG transport between samples A and B/C, which have an order of magnitude difference in the impurity levels at the GaN/AlN interface. The samples were measured from 300 K to 20 K, at 1 T magnetic field in a Van der Pauw configuration with soldered Indium corner contacts. The results are shown in \textbf{figure \ref{f3}}. Both samples show similar mobile hole densities $\sim5\times 10^{13}$ cm$^{-2}$ throughout measured the temperature range. These densities are expected from the difference in polarization at the AlN/GaN heterojunction. The holes do not freeze-out even at 20 K, confirming the 2DHG is highly degenerate and metallic in nature for both samples. 

A significant difference is observed in the measured Hall-effect mobilities. Sample A shows a room temperature mobility of $\sim5$ cm$^2/$Vs, whereas the sample B and C have a $2\times$ higher mobility of $\sim10 - 14$ cm$^2/$Vs. The samples show a similar trend of mobility as a function of temperature. This is explained by the freeze out of acoustic and optical phonons\cite{Bader}, and agrees with previous report\cite{Chaudhuri2019} of GaN/AlN 2DHG. However, the 2DHG mobility in sample B, C is consistently 2-3$\times$ higher that of sample A, with the highest mobilities of $85$ cm$^2/$Vs and $64$ cm$^2/$Vs respectively at 20 K. Other samples under similar growth conditions with IBLs have exhibited even higher mobilities, the highest\cite{Chaudhuri2019} among them is also plotted in figure \ref{f3} for reference. This boost in mobility of sample B and C results in $3\times$ lower sheet resistivity compared to sample A throughout the measured temperature range, as seen in \textbf{figure \ref{f3} (c)}. 

\begin{figure}[htb!]
\begin{center}
\includegraphics[keepaspectratio=true, width=0.475\textwidth]{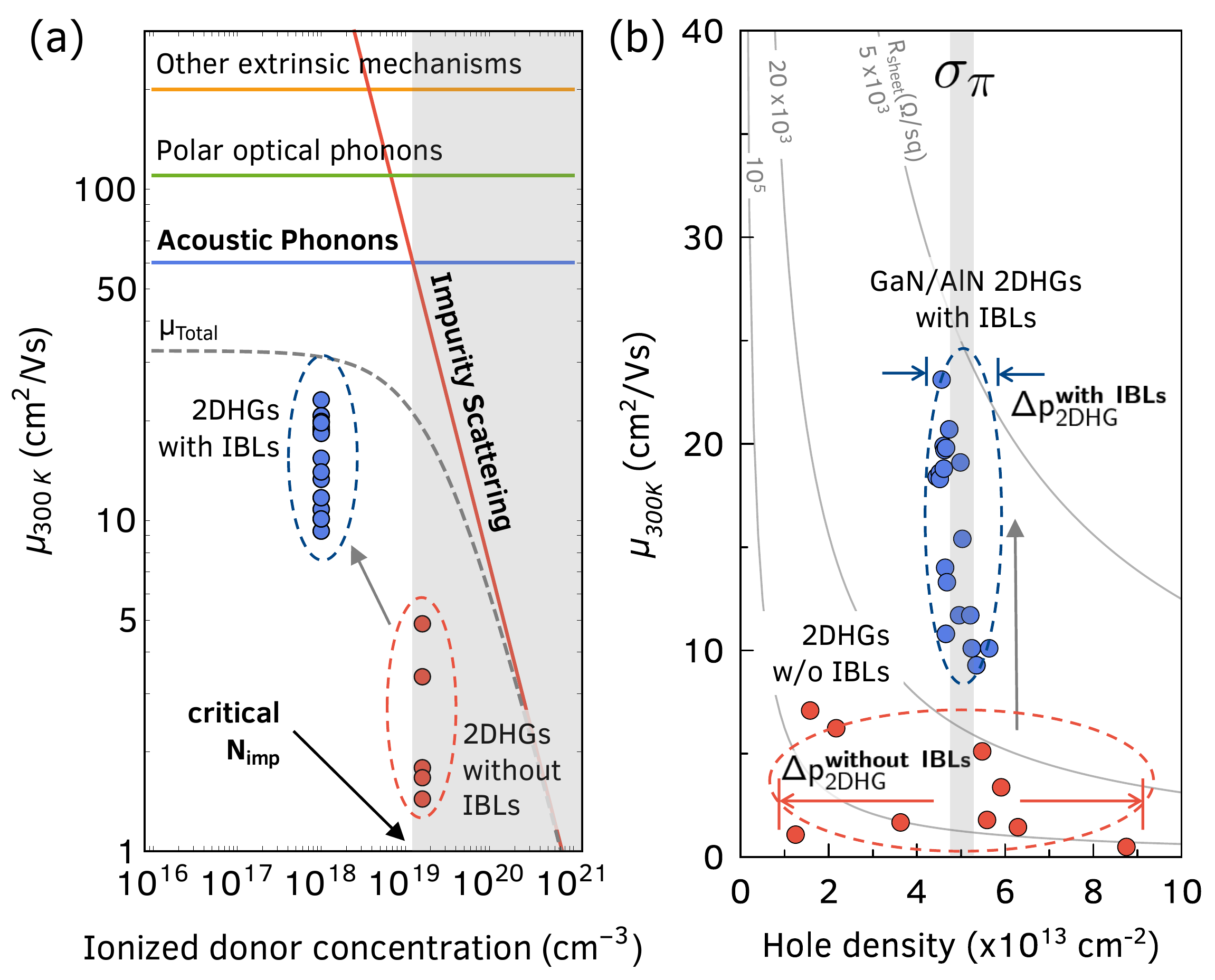}
\end{center}
\caption{\label{f4} (a) Theoretical room temperature (RT) mobility $\mu_\textrm{300K}$ of a GaN/AlN 2DHG of density $\sim$5$\times 10^{13}$ cm$^{-2}$, as a function of concentration of ionized background impurities $N_\textrm{imp}$. The 2DHG scattering is limited by ionized impurity scattering at impurity concentrations $N_\textrm{imp} \geq N_\textrm{imp}^\textrm{cr} \sim1.2 \times 10^{19}$ cm$^{-3}$. The 2DHGs with IBLs ensure that $N_\textrm{imp}^\textrm{cr}$ and thus have a higher phonon-limited mobility. (b) 300K mobility vs charge density of GaN/AlN 2DHG samples grown with and without the IBLs under similar conditions. A polarization-induced 2DHG density of $\sim5 \times 10^{13}$ cm$^{-2}$ is expected in all the samples. The samples without the IBLs, exhibit a larger spread in the 2DHG densities and lower mobilities. The samples with IBLs are in agreement to the expected 2DHG density and have higher mobilities, resulting in high-conductivity 2DHGs.}
\end{figure}

The comparison of SIMS provides clear evidence that the presence of charged donor impurities at the GaN-AlN interface plays an important role in limiting the mobility of the 2DHG in these structures. Quantitatively, the mobility of a 2DHG of density $p_s$ limited by Coulomb scattering from ionized background impurities of uniform 3D concentration $N_{\textrm{imp}}$ is given by\cite{Davies1997TheSemiconductors} : 
\begin{eqnarray}
\mu_{\textrm{imp}} \approx \frac{4 (2\pi)^{5/2} \hbar^3 (\epsilon_0 \epsilon_s)^2}{m_\textrm{eff}^2 \ e^3} \cdot \frac{p_s^{3/2}}{N_{\textrm{imp}}},
\label{eqn_IMP_scattering}
\end{eqnarray}
where $\hbar$ is the reduced Planck's constant, $e$ is the electron charge and $\epsilon_0$ is the vacuum permittivity. $\epsilon_s = 8.9$ is the relative dielectric constant of the GaN channel. This expression considers a simplified single parabolic valence band (VB) with hole effective mass \cite{Im1997RadiativeGaN} $m_\textrm{eff} = 2.0 \ m_0$. The value of the mobility for the GaN 2DHG of density $p_s = 5 \times 10^{13}$ cm$^{-2}$ due to ionized background impurities is plotted as a function of $N_{\textrm{imp}}$ in \textbf{figure \ref{f4}(a)}. It reduces from $\mu_{imp} \geq 100$ cm$^2$/V$\cdot$s at $N_{imp} \sim 5 \times 10^{18}$cm$^{-3}$ to $\mu_{imp} \sim 5$ cm$^2$/V$\cdot$s at $N_{imp} \sim 1.5 \times 10^{20}$cm$^{-3}$.

The dominant intrinsic scattering mechanism at room temperature for GaN/AlN 2D holes is due to acoustic phonons (AP) \cite{Bader} through the deformation potential coupling. For a simplified single valence band model, the AP-limited hole mobility of for a 2DHG at temperature $T$ is \cite{Davies1997TheSemiconductors} :
\begin{eqnarray}
\mu_{\textrm{AP}} \approx \frac{16 e \  \rho \  v_s^2 \  \hbar^3}{3 k_B T \  D^2 \ m_\textrm{eff}^2 \ b(p_s)} , 
\label{eqn_AP_scattering}
\end{eqnarray}
where $k_B$ is the Boltzmann constant. $b(p_s) = \left[ \left( 33 m_\textrm{eff} e^2 \ p_s \right) / \left( 8 \hbar^2 \epsilon_0 \epsilon_s \right) \right]^{1/3}$ is the variational Fang-Howard wavefunction parameter that quantifies the spatial spread of the 2DHG. Because the 2DHG is located in the GaN layer, the properties of GaN mass density $\rho = 6.15 \times 10^3$ kg/m$^3$ and sound velocity $v_s = 7963$ m/s are used with a scalar equivalent valence band deformation potential of magnitude $D = 6.2$ eV considered as a simplification of the deformation potential matrix for a single parabolic valence band approximation. The resulting calculated AP-limited 2DHG mobility using equation (\ref{eqn_AP_scattering}) agrees well with the numerically calculated values using multi-band $k\cdot p$ models\cite{Bader} over a wide range of temperatures. This $D$ value also concurs with the sets of GaN VB deformation potentials reported in the literature\cite{Yamaguchi1998DeterminationGaN, Ohtoshi1997DependenceLasers}. Using this simplified and calibrated model, the expected room-temperature AP-limited mobility of the GaN/AlN 2DHG is $\sim60 $ cm$^2$/Vs, which is plotted in \textbf{figure \ref{f4}(a)}.  

Other scattering mechanisms affecting the 2DHG that are less effective at room temperature are also included in \textbf{figure \ref{f4}(a)}. The room-temperature polar optical phonon (POP) limited mobility is set\cite{Bader} at 110 cm$^2$/Vs. An extrinsic scattering mechanism limited mobility, which encompasses the effect of mechanisms such as interface roughness, dislocations, etc., is set to 200 cm$^2$/Vs. Since the acoustic and optical phonon scattering modes are frozen out at low temperatures, the extrinsic value of 200 cm$^2$/Vs corresponds to the highest 10 K Hall mobility measured in GaN/AlN 2DHGs\cite{Chaudhuri2019}.  The resultant total mobility ($\mu_\textrm{Total}$) approximated by Matthiessen's rule is plotted as a function of ionized donor concentration in \textbf{figure \ref{f4}(a)}.

In a chemically and structurally pure GaN/AlN heterostructure in the low impurity limit, AP scattering is the dominant scattering mechanism for 2D holes at room temperature. However, it is evident from \textbf{figure \ref{f4}(a)} that there exists a critical background impurity concentration above which the ionized impurity scattering becomes the dominant scattering mechanism, and lowers the 2DHG mobility to below its intrinsic AP-scattering limit.  This critical impurity concentration $N_\textrm{imp}^\textrm{cr}$ is obtained by equating equations (\ref{eqn_IMP_scattering}) and (\ref{eqn_AP_scattering}): 
\begin{eqnarray}
N_\textrm{imp}^\textrm{cr} \approx A' \ T \ \frac{D^2 \ m_\textrm{eff}^{1/3} \  (\epsilon_0 \epsilon_s)^{5/3}}{\rho \ v_s^2} p_s 
\label{eqn_Nimp_critical}
\end{eqnarray}
where $A' = (120 \ k_B)/(\hbar^{2/3} \ e^{10/3})$ is a material independent constant. For convenient numerical estimates, equation (\ref{eqn_Nimp_critical}) is cast in the form : 
\begin{eqnarray}
N_\textrm{imp}^\textrm{cr} &\approx& 5 \times 10^{17} \ \textrm{cm}^{-3} \times \nonumber \\ & & \epsilon_s^{5/3} \left( \frac{D}{1 \  \textrm{eV}} \right)^2 \left( \frac{m_\textrm{eff}}{m_0} \right)^{1/3} \left( \frac{p_s}{10^{13} \  \textrm{cm}^{-2}} \right) \times \nonumber \\ & & \left( \frac{10^3 \  \textrm{kg/m}^3}{\rho} \right)  \left( \frac{10^3 \  \textrm{cm/s}}{v_s} \right)^{2} ,
\label{eqn_Nimp_critical_simpl}
\end{eqnarray}
where $m_0$ is the free-electron mass. Substituting the values for the GaN/AlN 2DHG system, a $N_\textrm{imp}^\textrm{cr} \sim1.2 \times 10^{19}$ cm$^{-3}$ is obtained. Thus it is expected that when the background impurity density exceeds $\sim1.2 \times 10^{19}$ cm$^{-3}$, impurity scattering is dominant and the resultant 2DHG mobility is lower than the intrinsic AP-limited mobility.

To compare with this theoretical prediction, the experimentally measured room-temperature Hall-effect mobilities of the GaN/AlN 2DHGs with AlN buffer layers grown without IBLs (similar to sample A) and with IBLs (similar to sample C) are plotted in \textbf{figure \ref{f4}(a)}.  They fall in two blocks.  The impurity densities at the 2DHG interface are estimated to be 1.6$\times$10$^{19}$ cm$^{-3}$ for the former and 10$^{18}$ cm$^{-3}$ for the latter, following the SIMS results of sample A and C. The experimental data agree with the trend predicted by the model: the 2DHGs with $N_\textrm{imp} \geq N_\textrm{imp}^\textrm{cr}$ have a lower mobility than those with $N_\textrm{imp} \leq N_\textrm{imp}^\textrm{cr}$, supporting the hypothesis that impurity scattering plays a dominant role in transport of 2DHGs without IBLs. Furthermore, since equation (\ref{eqn_Nimp_critical_simpl}) is a function of basic material parameters, it serves as a useful rule-of-thumb for designing and optimizing heterostructures with 2D carriers not limited to the GaN/AlN 2DHG studied here. The deviation from theory in \textbf{figure \ref{f4}(a)} is attributed to the single parabolic VB assumption for the mobility calculations, which ignores inter-valence band scattering. The low mobility samples could possibly also have impurity concentrations higher than 1.6$\times$10$^{19}$ cm$^{-3}$ leading to the spread in their mobilities. 

If the donor impurity concentration is high enough that its integrated sheet concentration in GaN becomes comparable to the mobile hole density of $\sim5 \times 10^{13}$ cm$^{-2}$, the donors can compensate the holes resulting in significant deviation from the expected hole density $\sigma_{\pi} \sim5 \times 10^{13}$ cm$^{-2}$.  This will also affect the mobility of the uncompensated holes due to neutral impurity scattering.  The strong effect on the hole density is clearly seen in \textbf{figure \ref{f4}(b)}, where the room temperature 2DHG mobilities versus densities are plotted for samples with and without IBLs. The samples without IBLs show a significantly large spread in the 2DHG densities with $\Delta p_s \sim 7 \times 10^{13}$ cm$^{-2}$ and low mobilities.  The 2DHG in the samples with IBLs consistently exhibit a higher mobility, and densities as expected of $\sim5 \times 10^{13}$ cm$^{-2}$ in all samples. More importantly, they exhibit tight control over the density which shows a 7$\times$ reduction in the spread to $\Delta p_s < 1 \times 10^{13}$ cm$^{-2}$. The variation in the 2DHG mobilities among these samples with IBLs is believed to be due to other non-dominant extrinsic scattering mechanisms such as interface roughness scattering or dislocation scattering. Varying the extrinsic scattering mechanism limited mobility from 200 cm$^2$/Vs to 60 cm$^2$/Vs is expected to cause a $\sim10$ cm$^2$/Vs variation of mobilities at room temperatures and $\sim100$ cm$^2$/Vs at low temperatures. These are in agreement with the measured mobility data spread observed in \textbf{figures \ref{f3}} and \textbf{\ref{f4}}. 

Thus, suppressing the impurities using IBLs not only results in higher mobilities, it also enables tightly controlled 2DHGs densities. An AlN buffer layer with 4 IBLs or a combination of N-rich NL and 1 IBL is efficient enough to block the impurities from affecting the 2DHG. As a further confirmation of the effectiveness of the IBLs, a UID-GaN/AlN 2DHG sample with a thinner $\sim100$ nm AlN buffer layer with N-rich NL and only 1 IBL showed a room temperature mobility of $\sim10$ cm$^2/$Vs at density of $4.4\times 10^{13}$ cm$^{-2}$. These optimized GaN/AlN 2DHGs offer the highest room temperature p-type conductivity compared to other single-channel 2DHGs reported in III-nitrides\cite{Chaudhuri2019}.

\section{Conclusions}

This study thus has taken an important step towards answering why the 2DHG in undoped GaN/AlN structures remained elusive nearly 2 decades after its prediction and the experimental discovery and subsequent technological impact of its n-type dual, the Al(Ga)N/GaN 2DEG.  As has been the case in semiconductor physics\cite{Orton2008TheSemiconductors}, the control of impurities during material growth plays a decisive role in all semiconductors, and especially in wide-bandgap semiconductors. For example, the presence of Hydrogen limited the conductivity of MOCVD grown Mg-doped GaN for decades, before its effect was discovered, understood and solved\cite{Nakamura2015}. Driven by the commercialization of GaN RF HEMTs, the growth of GaN buffer layers has been studied and optimized  on various substrates for high-quality Al(Ga)N/GaN 2DEGs. Taking advantage of this established platform, most of the earlier 2DHGs reported have been on GaN buffer layers, in metal-polar AlInGaN/GaN \cite{Hahn2013}, InGaN/GaN \cite{Zhang2016}, GaN/AlGaN/GaN \cite{Nakajima2010} structures. 

Only in 2010s has the availability of AlN as a photonic and electronic device platform led to the exploration of AlN as a buffer layer for growing 2DEGs\cite{Qi2016, Li2010} and 2DHGs. In 2013, the first 2DHG in a GaN/AlN heterostructure with Mg-doped GaN layer was reported\cite{Li2013}. These films had low 2DHG mobilities, similar to that in sample A in this report, and unrepeatable 2DHG densities across growths. A Mg-doped GaN cap layer was necessary to stabilize the 2DHG. Since then, further refinements of the AlN buffer layer growth summarized in this work have led to increase in 2DHG mobilities and the observation of high-conductivity 2DHG in an undoped GaN/AlN heterostructure. The authors would like to highlight that measuring a buried 2DHG in an undoped wide bandgap heterostructure is inherently difficult due to the inability to make direct ohmic contacts to carriers. Hence the lack of reports of undoped GaN 2DHGs in literature might be partly due to the inability to detect and measure high-resistivity undoped buried 2DHGs.

To summarize, this report shows that for metal-rich MBE growth, surface segregation of Si and O impurities significantly degrades the density control and transport properties of 2DHGs at undoped GaN/AlN heterostructures.  A combination of N-rich nucleation layer and IBLs suppresses this undesired effect, boosting the 2DHG mobilities and resulting in a tighter control over the 2DHG density. Though this work was for MBE grown structures on AlN on sapphire templates, the same principles apply to MOCVD growths and to AlN films grown on other substrates (e.g. Silicon, SiC, single-crystal bulk AlN etc). The findings reported here are also important for UV photonic devices where p-type layers are grown on the top of Si-doped n-layers. Similar precautions must be taken to avoid undesired compensation of the very low hole densities and reduction in their mobilities. The same issues are also expected to arise if the polarity of the films are flipped, and a 2DEG is generated at the GaN/AlN interface on N-polar AlN buffer layer. With the rising relevance of AlN as the platform for UV photonics and future RF electronics\cite{Bader2020ProspectsDevices, Hickman2021NextPlatform}, significant interest exists in using 2DEGs and 2DHGs on AlN to make RF p-channel\cite{Bader2019e, nomotoIEDM2020} and n-channel transistors\cite{Hickman2019, Hickman2020FirstGHz} and enable wide-bandgap RF CMOS type devices. Combining the results of this work with recent advancements in homoepitaxial growths of AlN\cite{Lee2020, Cho2020MolecularCleaning} should enable fundamental scientific studies of 2DEGs and 2DHGs in such polar heterostructures and simultaneously enable significant technological advances.

\section*{Supplementary Material}
See supplementary material for details of the molecular beam epitaxial growth of the heterostructures studied in this work.

\begin{acknowledgments}
This work was supported by Intel Corp., AFOSR (grant FA9550-20-1-0148) and NSF (grants 1710298, 1534303). Characterizations and measurements were carried out in part at CNF, supported by NSF Grant NNCI-2025233, and at CCMR Shared Facilities, supported through the NSF MRSEC program (DMR-1719875) and NSF MRI (DMR-1429155 and DMR-1338010) programs. Z.C. was funded through PARADIM as part of the NSF Materials Innovation Platform program (DMR-1539918). 
\end{acknowledgments}

\section*{Data Availability}
The data that supports the findings of this study are available within the article and its supplementary material. 

\section*{References}

\bibliography{ms.bib}

\end{document}



\title{{\large Supplementary Material for} \\ High conductivity Polarization-induced 2D hole gases in Undoped GaN/AlN Heterojunctions enabled by Impurity Blocking Layers} 

\author{Reet Chaudhuri}
\email[]{rtc77@cornell.edu}

\affiliation{School of Electrical and Computer Engineering, Cornell University}

\author{Zhen Chen}
\affiliation{Department of Applied and Engineering Physics, Cornell University}

\author{David Muller}
\affiliation{Department of Applied and Engineering Physics, Cornell University}
\affiliation{Kavli Institute of Nanoscience, Cornell University}

\author{Huili Grace Xing}
\affiliation{School of Electrical and Computer Engineering, Cornell University}
\affiliation{Department of Material Science and Engineering, Cornell University}
\affiliation{Kavli Institute of Nanoscience, Cornell University}

\author{Debdeep Jena}
\affiliation{School of Electrical and Computer Engineering, Cornell University}
\affiliation{Department of Material Science and Engineering, Cornell University}
\affiliation{Kavli Institute of Nanoscience, Cornell University}

\date{\today}

\pacs{}

\maketitle 

\section{MBE Growth recipes}

The heterostructures studied in this work were grown in a Veeco Gen10 plasma assisted molecular beam epitaxy (PA-MBE) system, with standard effusion sources for Ga, Al and an RF plasma source for N. 8mm $\times$ 8mm diced pieces of metal-polar MOCVD-grown AlN on sapphire templates from DOWA were used as starting substrates. The substrates were ultrasonicated in acetone, methanol and iso-propanol for 15 minutes each and mounted on lapped 3-inch Si wafers using molten indium. They were then out-gassed at 200$^{\circ}$C for 7 hours and 500$^{\circ}$C for 2 hours under vacuum before introducing into the MBE growth chamber. For the active nitrogen flux during growth, the RF plasma power was 400 W and the N gas flow rate was set at 2 sccm, corresponding to a calibrated N-limited growth rate of $0.52 \ \mu$m/hour. Magnesium and silicon dopant cells were kept cold throughout the growth. Reflection high energy electron diffraction (RHEED) was used for in-situ monitoring of the growth. The growth chamber pressure was $\sim2.01\times10^{-5}$ Torr during the growth.

The growth recipe for \textbf{sample A} is illustrated in figure \ref{fS1}. The sample was heated to T$_{\text{sub}}$ = 780$^{\circ}$C for AlN buffer growth. The first $\sim$30 nm of AlN, the nucleation layer (NL), was grown under N-rich conditions by opening both Al and N shutter simultaneously at Al BEP of $\sim6.1\times10^{-7}$ Torr, corresponding to an Al:N flux ratio $\sim$0.9. The Al cell temperature was then ramped up at 3$^{\circ}$C/min to a final BEP flux of $\sim9.1\times10^{-7}$. While the cell temperature ramped up, migration enhanced epitaxy (MEE) cycles were performed to smoothen out the growth surface. The Al shutter and N shutter were alternatively opened for 30 s each. As the Al flux slowly ramped up, the monitored RHEED pattern went from spotty, indicating a rough surface expected from a N-rich AlN growth\cite{Koblmueller2003}, to streaky, indicating a smooth surface with a coalesced film. Once the Al cell was at the targeted cell temperature corresponding to BEP flux of $\sim9.1\times10^{-7}$, the rest of the buffer was grown under metal-rich AlN condition by opening both the Al and N shutters simultaneously. Since Al desorption rate from AlN growth surface is low at T$_{\text{sub}}$ = 780$^{\circ}$C, the accumulated Al metal was periodically consumed after 5 mins of AlN growth by closing the the Al shutter and keeping the N shutter open. The consumption of Al was monitored by tracking the saturation of the RHEED pattern intensity\cite{Chaudhuri2020}. A total of $\sim420$ nm thick AlN buffer was grown in this manner. This technique of consumption of excess Al on the surface also ensures a sharp, abrupt AlN-GaN interface\cite{Chaudhuri2020}, minimizing the interface roughness scatterring of the 2D carriers. The substrate was then cooled down to T$_{\text{sub}}$ = 750$^{\circ}$C for the UID-GaN growth. The Ga shutter was opened for 10 s at a Ga BEP flux of $\sim1\times 10^{-6}$ Torr before the N shutter was opened to wet the growth surface, following which the N shutter was also opened to grow the 15nm thick UID-GaN channel.

Figure \ref{fS2} shows the growth recipe followed for \textbf{sample B}. The AlN buffer was grown at T$_{\text{sub}}$ = 780$^{\circ}$C. The AlN was grown in metal-rich conditions in a 3-step cycle consisting of (1) Ga pre-deposition, (2) Al-rich AlN growth and (3) N anneal. Ga BEP flux of $\sim1\times 10^{-6}$ Torr was initially deposited on the surface for $\sim$30 s while N and Al shutter remained closed. Both the Al and N shutter were then opened simultaneously to grow AlN under metal rich conditions. After 5 min of AlN growth, the Al shutter was closed and the excess Al metal on the surface was consumed by the incident N flux\cite{Chaudhuri2020}. The pre-deposited Ga has a dual role here - to act as a surfactant for the subsequent AlN growth\cite{Okumura2010Enhancement0001} and to form the high-composition AlGaN IBLs. A small amount of this pre-deposited Ga metal incorporates into the initial layers of the AlN crystal to form high-composition $>90 - 95\%$ AlGaN, due to non-idealities of the growth conditions - N leak from around the shutter when it is closed and instantaneous variance in Al and N fluxes when the shutter is opened. Reports have also shown that a small amount of Ga is incorporated into AlN crystal when used a surfactant for AlN growth\cite{Mayboroda2017GrowthGrowth}. The the AlGaN composition and thickness is controlled by the Ga BEP flux, deposition time and substrate temperature during the AlN growth, which were optimized separately. In sample B, each AlN cycle resulted in 1 IBL and $\sim55$ nm of AlN layer growth. Repeated 8 times, this corresponds to $\sim430$ nm of AlN buffer and 8 IBLs in the buffer. The substrate was then cooled down to T$_{\text{sub}}$ = 750$^{\circ}$C and the 15 nm of UID-GaN layer was grown on top of the AlN buffer.

Figure \ref{fS3} shows the growth recipe followed for sample \textbf{sample C}. It combined the growth techniques of samples A and B to incorporate an N-rich AlN nucleation layer as well as periodic IBLs. A $\sim$30 nm of N-rich AlN was grown at T$_{\text{sub}}$ = 780$^{\circ}$C with Al flux of $\sim6.1\times10^{-7}$ Torr. This was followed by 30s + 30s MEE cycles of alternating Al and N fluxes while the Al cell was heated up. After the Al flux of $\sim9.1\times10^{-7}$ Torr is reached, 5 periodic cycles of Al-rich AlN with 30 s Ga pre-deposition were grown, in conditions similar to sample B. The total AlN buffer thickness Was $\sim440$ nm with 5 IBLs separated by $\sim55$ nm. After the AlN growth, the substrate was cooled down to T$_{\text{sub}}$ = 750$^{\circ}$C and the 15nm UID-GaN layer was then grown on top of the buffer. 

Figure \ref{fS3} also includes representative RHEED patterns from the AlN crystal surface from the same orientation at different stages of the growth. The RHEED pattern starts off streaky before growth, signifying a smooth starting AlN template surface. The RHEED pattern turns spotty after the N-rich AlN nucleation, indicating the roughening and incomplete coverage of the AlN surface, expected from a N-rich AlN growth \cite{Koblmueller2003}. RHEED turns streaky after the MEE cycles. Also shown is a comparison of the RHEED pattern after 5 min of Al-rich AlN growth and after the subsequent consumption of excess Al by nitrogen. The RHEED pattern remains the same, but the streaks become brighter and saturate, indicating the absence of Al metal on the surface\cite{Chaudhuri2020}.

\bibliography{supplement.bib}

\newpage

\renewcommand{\thefigure}{S\arabic{figure}}

\begin{figure*}[h]
\begin{center}
\includegraphics[keepaspectratio=true, width=\textwidth]{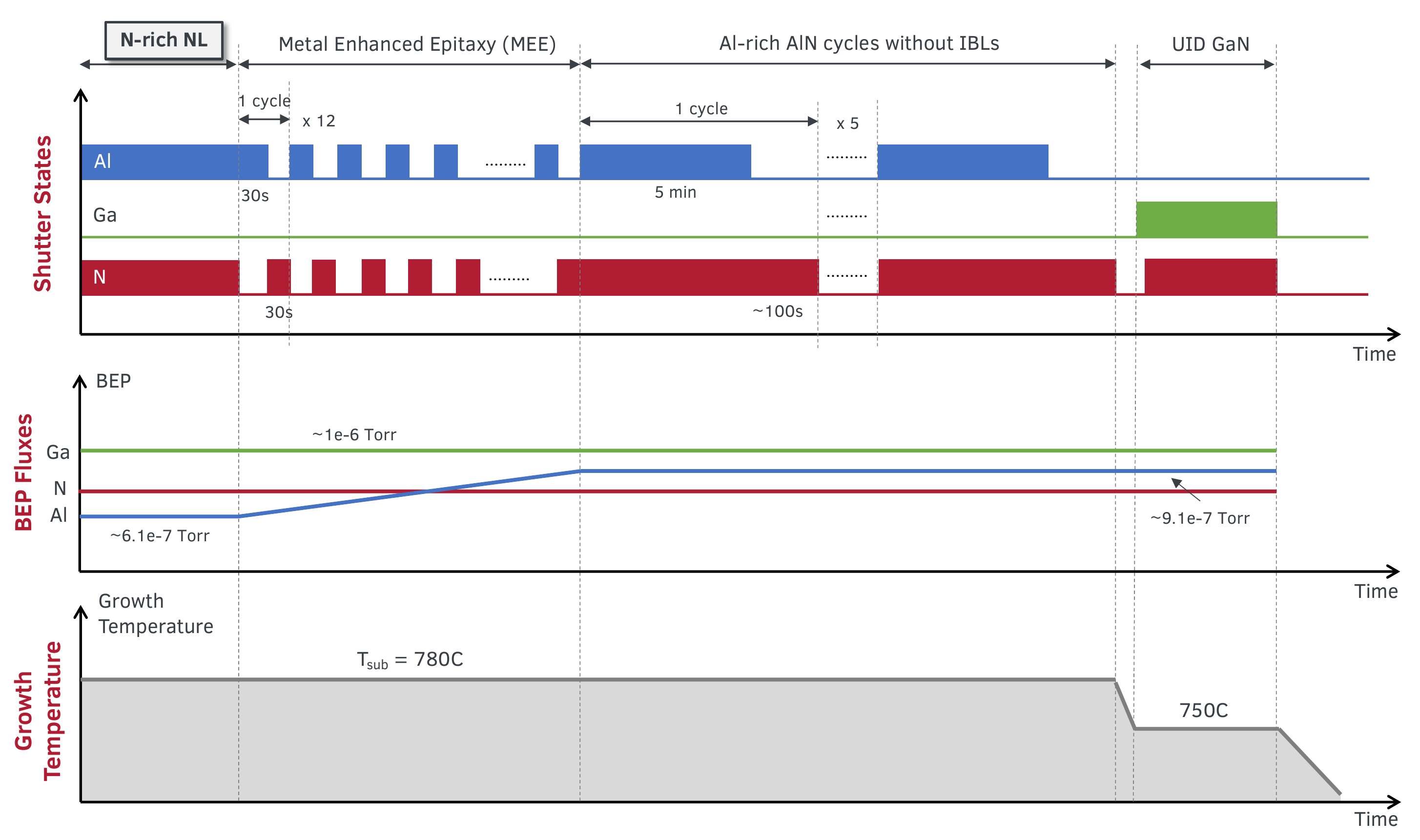}
\end{center}
\caption{\label{fS1} (a) Growth diagram sample A, with nucleation layer grown under N-rich conditions by maintaining an Al:N flux ratio $< 1$}
\end{figure*}

\begin{figure*}[h]
\begin{center}
\includegraphics[keepaspectratio=true, width=\textwidth]{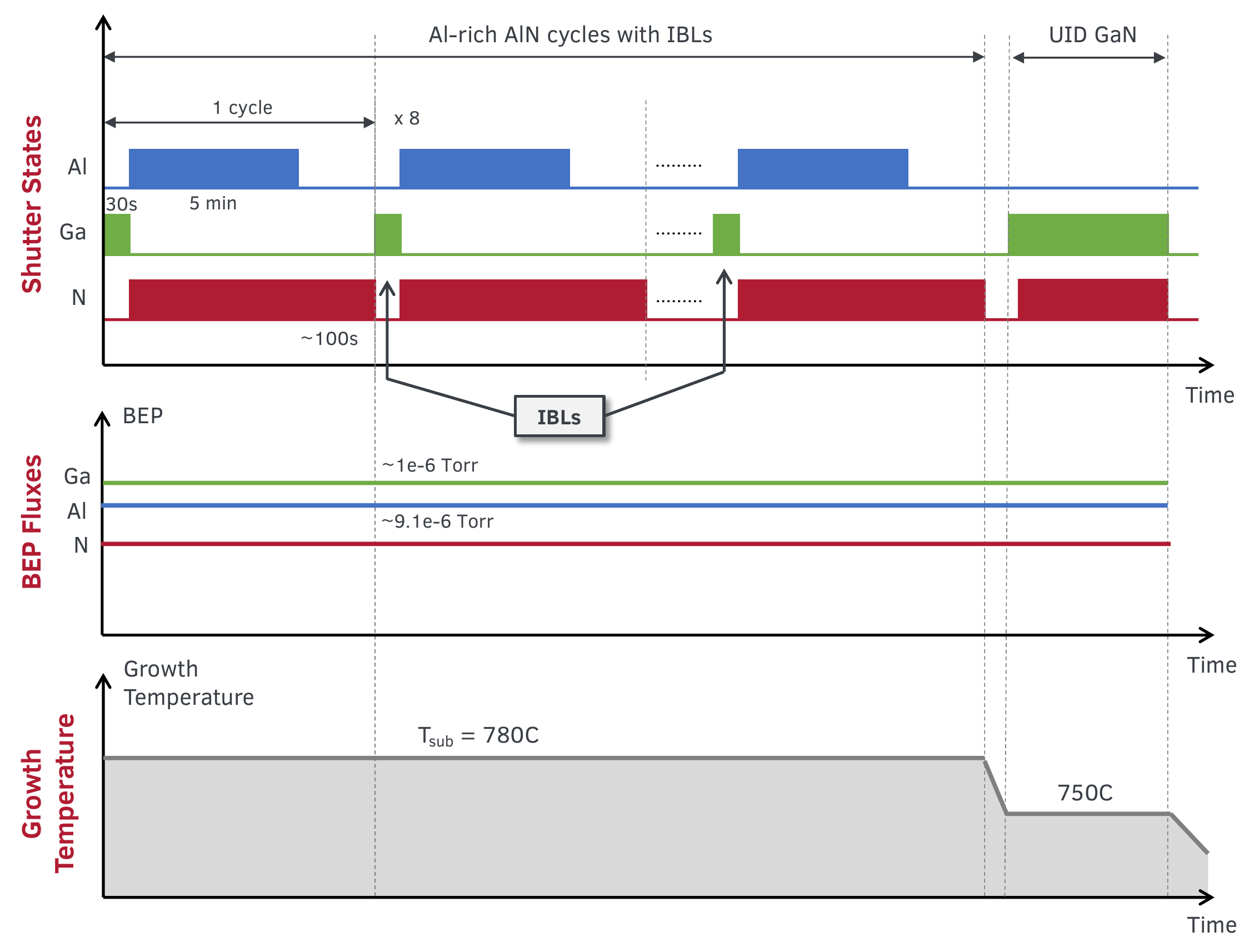}
\end{center}
\caption{\label{fS2} (a) Growth diagram for sample B, with impurity blocking layers (IBLs) incorporated in the AlN buffer grown under Al-rich growth conditions.}
\end{figure*}

\begin{figure*}[h]
\begin{center}
\includegraphics[keepaspectratio=true, width=\textwidth]{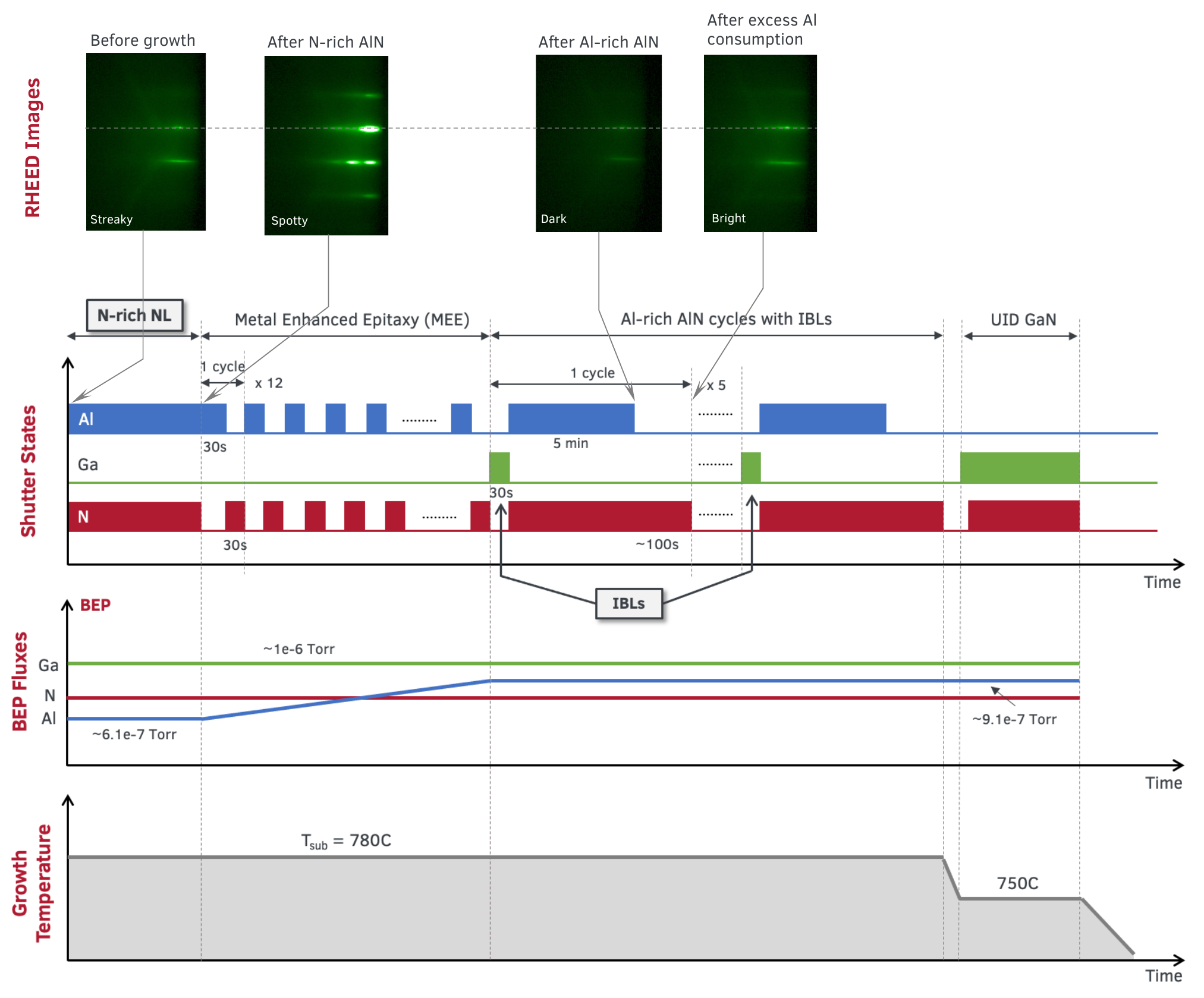}
\end{center}
\caption{\label{fS3} (a) Growth diagram for sample C, with both N-rich nucleation of AlN and IBLs in the buffer. Characteristic RHEED pattern evolution at various stages of the growth is also shown.}
\end{figure*}